\documentclass[aps,pre,twocolumn,floats,floatfix,nofootinbib,superscriptaddress]{revtex4-1} % ,eqsecnum removed
\usepackage{amsmath}
\makeatletter

\renewcommand*{\p@subsection}{}

\renewcommand*{\p@subsubsection}{}
\makeatother

\usepackage{natbib}
\setcitestyle{square}

\usepackage{graphicx}
\usepackage{sidecap}

\usepackage[usenames,dvipsnames]{xcolor}
\usepackage[colorlinks=true,urlcolor=Blue,citecolor=Violet,linkcolor=BrickRed]{hyperref}

\newcommand{\be}{\begin{equation}}
\newcommand{\ee}{\end{equation}}

\newcommand{\fig}[1]{Fig.\,\ref{#1}}

\newcommand{\eq}[1]{Eq.\,(\ref{#1})}
\newcommand{\Eq}[1]{Equation\,(\ref{#1})}

\newcommand{\eqsand}[2]{Eqs.\,(\ref{#1}) and (\ref{#2})}

\def\r{\mathbf{r}}
\def\k{\mathbf{k}}
\def\B{\mathbf{B}}

\def\d{\mathrm{d}}
\def\Y{{\cal Y}} % for compatibility with previous publications, but 4\pi included

\begin{document}

\title{Larmor Frequency in Heterogeneous Media}
\author{Valerij G.\ Kiselev}
\affiliation{Medical Physics, Dept.\ of Radiology, Faculty of Medicine, University of Freiburg, Germany}
\date{\today}

\begin{abstract}
\noindent  
The Larmor frequency shift is found in porous media consisting of NMR-reporting fluid filling a connected pore within an NMR-invisible matrix for the case of fast diffusion in the fluid. The matrix material has a distinct location-independent anisotropic magnetic susceptibility tensor that induces a heterogeneous microscopic magnetic field when exposed to the strong main field of an NMR device. Aside from the connectivity of the pore, the matrix geometry is arbitrary. 

%\makebox[0pt]{\raisebox{180pt}[0pt][0pt]{\hspace{1.62\textwidth} {\color{red} Draft; do not distribute}}}

\end{abstract}

\maketitle

%%%%%%%%%%%%%%%%%%%%%%%%%%%%%%%%%%%%%%%%

\section{Introduction}
 
\begin{figure}[!b]
\includegraphics[width=\columnwidth]{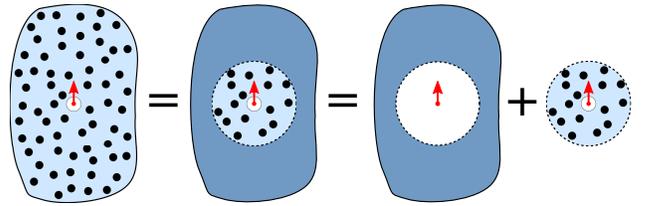}
\caption{A sample with a microscopic magnetic structure and the decomposition of its magnetization for calculating the magnetic field at the position of an NMR-reporting spin (red arrow). The heterogeneous magnetization is shown as inclusions (black dots) with a magnetic susceptibility different from that of the surrounding fluid (light blue). The first equality: The sample can be divided into a far and a spherical near region, the {\em mesoscopic sphere} (the sphere size is greatly exaggerated in the image). The far region contributes a field proportional to its average (homogenized) susceptibility (darker blue). The second equality: The field at the spin's position is the sum of two contributions induced by the far region with excluded mesoscopic sphere, ${\bf B}^{\rm macro}$, in \eq{2fields}, and the mesoscopic sphere. The field from within the sphere, ${\B}^{\rm meso}$ in \eq{2fields}, needs to be calculated given the medium microarchitecture. The small sphere stands for the classical Lorentz sphere in the NMR-reporting fluid.}
\label{fig:sample_decomposition}
\end{figure}

Precise measurement of the signal phase in the human brain at high magnetic field \cite{Abduljalil2003,Duyn2007,Marques2009} reanimated the interest in calculating the Larmor frequency shift in media with heterogeneous magnetic properties, in particular brain white matter with the account for the distinct magnetic susceptibility of myelin. In the focus of intensive discussion, this problem has decoupled from its original biomedical context. From the physics point of view, the challenge is to calculate the measurable phase of NMR signal acquired in a medium consisting of an NMR-reporting liquid (referred to as water in what follows) and numerous microscopic inclusions with a magnetic susceptibility different from that of water, \fig{fig:sample_decomposition}. 

To date, it is understood that the contribution of the local environment in the precession frequency of NMR-reporting spins is linked to the microscopic magnetic architecture of the medium, which might be too complex to be described by the classical Lorentz sphere construction \cite{He2009,Yablonskiy2014}; according to the currently available theory, magnetized inclusions of different types contribute {\em additively} to the Larmor frequency offset \cite{Sukstanskii2014,Yablonskiy2015,Yablonskiy2017}, whereas each type involves the Lorentz cavity construction with a specific shape; the shape is selected to mimic the true microscopically calculated frequency shift. While this works for inclusions of the simplest shapes of sphere and cylinder, in general, the Lorentz cavity construction has been abandoned for direct calculation of the frequency shift with account for tissue microarchitecture \cite{Sukstanskii2014,Wharton2015,Yablonskiy2015,Yablonskiy2017}. The additive contribution of different inclusions has been recognized as the limit of their low volume fraction \cite{Duyn2014}, which is a serious confounding factor for biological application.  

In this work, I develop a theory for the case of fast water diffusion in connected space outside magnetized inclusions of {\em arbitrary density and geometry}. Other inclusions' properties are treated in the simplest possible way. Namely, the inclusions are NMR-invisible and have all the same {\em position-independent} anisotropic magnetic susceptibility. A way to more realistic models of white matter is discussed at the end of this paper. The main result is a formula for the Larmor frequency offset in which the medium's magnetic microarchitecture enters via the {\em correlation function} of magnetized inclusions, \eq{meanfield} below; this formula generalizes the recent result for the case of isotropic magnetic susceptibility \cite{Ruh2018}. In the most general case, the frequency offset depends on five relevant parameters of the correlation function and does not depend on an infinite set of parameters defining its full functional form. In the case of axially symmetric media, such as nerve tissue studied ex-vivo \cite{Luo2014,Lee2010,Wharton2015}, there is a single relevant  parameter representing the medium microarchitecture, \eq{meanfield_ax_meso} below. This result is applied to quantify an empiric microstructural parameter introduced to interpret the frequency offset in excised optic nerve \cite{Wharton2015}. 

\section{Setting the scene}

The above described medium belongs to the class of porous media and the corresponding terminology is used hereafter. Consider a macroscopic sample that consists of a water-filled connected pore and an NMR-invisible matrix occupying the fraction $\zeta$ of the sample volume. The matrix is described by an indicator function $v(\r)$ that is unity inside the matrix and zero in the pore. The matrix material is characterised by anisotropic magnetic susceptibility described by a tensor $\chi_{ab}$, where $a,b\dots$ label the three spatial components of vectors and tensors. To simplify equations, the magnetic susceptibility is considered relative to water, which means that any measured Larmor frequency implies subtraction of the frequency measured in a sample of the same shape, but without the matrix. The matrix material is considered to be non-ferromagnetic, which means that $|\chi_{ab}|\ll 1$. This condition enable finding the macroscopic matrix magnetization as 
\be\label{M=}
M_a = \chi_{ab} B_{0,b} \,,
\ee
where $B_{0,b}$ is the $b$-component of the main field, $\B_0$. The additional macroscopic magnetic field induced by the matrix takes the form  
\begin{equation} \label{DB} 
\Delta B_a(\r)  = \int \d^3r_0 \, \Y_{ab}(\r-\r_0) \, M_b(\r_0) \, ,
\end{equation}
where the Einstein convention about the summation over repeated indices is used. $\Y_{ab}$ is the elementary dipole field, 
\begin{equation} \label{Dab0}
\Y_{ab}(\r) = \frac{3 \, \hat r_a \hat r_b -\delta_{ab}}{r^3} \,,
\end{equation}
where the hat denotes unit vectors, $\hat r_a=r_a/r$ and $\delta_{ab}$ is the Kronecker delta, which is unity for $a=b$ and zero otherwise. Since \eq{Dab0} has the form of a convolution, the field can be efficiently calculated in the Fourier domain, 
\begin{equation}\label{DBk}
\Delta B_a(\k)  = \Y_{ab}(\k) M_b(\k) \,.
\end{equation}
Note that the same root letters are used throughout this paper for the original and Fourier-trans\-formed quantities; the argument is always given explicitly to avoid confusion. The Fourier transform of the dipole field, \eq{Dab0}, takes the following form in the cgs system \cite{Salomir2003,Marques2008}:
\begin{equation}\label{DabLk0}
\Y_{ab}(\k) = 4\pi \left( \frac{\delta_{ab}}{3} - \frac{k_a k_b}{k^2}\right) \,.
\end{equation}
This expression takes into account the field of the Lorentz sphere for water molecules, $8\pi \delta_{ab}/3$ \cite{Dickinson51,Chu90}, which is discussed in more detail below.   

The time-averaged magnetic field experienced by a water proton does not coincide with the value given by \eq{DB} because the condition of averaging over the molecular scale does not fulfill for the closest environment of the spin. Following the original idea of Lorentz \cite{Lorentz1880}, the effect of this environment should be considered with account for its microscopic structure and dynamics. In isotropic liquids such as water, this leads to the zero on average field from the nearest environment, which is taken effectively into account by using the macroscopic \eq{DB} with an infinitesimal spherical cavity, the Lorentz sphere \cite{Dickinson51,Chu90}. 

In media with microscopic structure, many orders of magnitude coarser than the molecular dimensions, \eq{DB} is valid on the microscopic scale. However, the effect of heterogeneous field should be specifically averaged to yield the overall Larmor frequency from  the whole macroscopic sample (or an MRI voxel). This averaging occurs on the scale much coarser than the microscopic one, but much finer than the macroscopic sample size. Such a scale is called {\em mesoscopic} in physics. Performing the necessary averaging on this scale is the overarching goal of the present study. 

This is achieved by replicating the original idea of Lorentz on the mesoscopic scale \cite{Wolber2000,Durrant2003}: The whole sample is subdivided in a near and a far regions relative to any considered NMR-reporting spin. While the far region contributes the field according to the macroscopically averaged medium parameters, the field in the near region should be calculated with account for the medium microscopic structure and the diffusive motion of water molecules. It is convenient to select the near region in the form of a sphere, \fig{fig:sample_decomposition}. This sphere is called the {\em mesoscopic sphere} in what follows. Its size is much smaller than the sample dimensions, but large enough to enable a smooth transition from the local environment of a given proton to the macroscopically averaged medium properties. In other words, the sphere size is large enough to include a statistically representative portion of the medium. In this study, I consider the case of fast diffusion, which means that the diffusion length in each direction is very large on the microscopic scale for the typical time of the signal acquisition, still it should be well below the size of the mesoscopic sphere. Deviations from this condition for isotropic media are discussed elsewhere \cite{Ruh2018}. 

Accordingly to the sample decomposition, the averaged deviation from the main field, $\Delta \bar\B$, experienced by a water proton consists of two terms, 
\begin{equation} \label{2fields}
\Delta \bar\B = \B^{\rm macro} + \B^{\rm meso}  \,,
\end{equation}
where ${\bf B}_{\rm macro}$ is the field created by a macroscopically homogeneous sample of the given shape with a small spherical cavity in the place of the mesoscopic sphere around the reporting spin and ${\bf B}_{\rm meso}$ is the field of the mesoscopic sphere, \fig{fig:sample_decomposition}. Finding the latter is the goal of the subsequent calculations.

\section{Frequency offset inside mesoscopic sphere}
\label{sec:meanfield}

The local Larmor frequency offset within the mesoscopic sphere is given by the longitudinal projection of the susceptibility induced magnetic field, 
\begin{equation}\label{Omega(r)}
\Omega(\r) = \gamma n_a  B^{\rm meso}_a(\r) \,,
\end{equation}
where ${\bf n} = \B_0 /B_0$ is the unit vector in the direction of the main magnetic field. This field varies over the characteristic length defined by the medium structure. The case of fast diffusion considered here implies that during the measurement the typical spin samples a significant portion of the medium. In this case, called the diffusion or motional narrowing, the reported Larmor frequency, $\overline \Omega$, is defined by the {\em spatial averaging} of the local frequency, $\Omega(\r)$, over the pore space,  
\begin{equation}
\overline \Omega =  \int \frac{\d^3 r}{(1-\zeta)V} \, \Omega(\r) \, [1-v(\r)] \, .
\label{meanfield0} 
\end{equation} 
While the integration here is performed over the whole volume of mesoscopic sphere, $V$, the matrix volume is excluded by the indicator function $1-v(\r)$ with the denominator, $(1-\zeta) V$, written for the normalization on the pore volume. 

\begin{figure}[tpb]
\includegraphics[width=0.6\columnwidth]{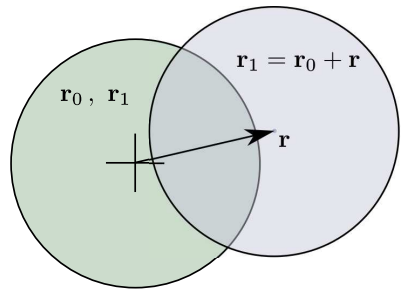}
\caption{Illustration of the integration variable change on the transition from  \eq{meanfield1} to \eq{meanfield2}. The original integration over $\r_0$ and $\r_1$ spans the sphere centered around the origin (the cross). The substitution $\r_1 = \r_0 +\r$ shifts this sphere by the vector $\r$, but the original limits on the $\r_1$ integration reduces the $\r_0$ integration volume to the intersection of the two spheres. This volume is large enough to provide for the averaging that gives the correlation function, \eq{def1Gamma}.  This condition is violated for large $\r$ for which the intersection collapses, but such values are not relevant because $\Gamma(\r)$ is essentially nonzero only for $r$ much smaller than the sphere size. }
\label{fig:spheres}
\end{figure}

The field within the sphere, $ \Omega(\r)$,\,  is found according to \eq{DB} with the magnetization from \eq{M=}. This gives  
\begin{widetext}
\begin{equation}\label{meanfield1} 
\frac{\overline \Omega\phantom{_0}}{\Omega_0} =   \int \frac{\d^3 \r_1}{(1-\zeta) V} \, \d^3 \r_0 \,  [1-v(\r_1)] \,n_a\Y_{ab}(\r_1-\r_0)\chi_{bc} n_c \,v(\r_0)\, ,
\end{equation} 
\end{widetext}
where $\Omega_0=\gamma B_0$ is the nominal Larmor frequency. The unity in the brackets gives rise to the integral over $\r_1$, which is proportional to the field induced by a homogeneously magnetized sphere. Since the elementary dipole field, \eq{DabLk0}, takes into account the field of the Lorentz sphere, the result is identically zero. In the remaining integral, which is bilinear in $v(\r)$, the variable $\r_1$ is substituted with $\r_0+\r$, which gives 
\begin{equation}\label{meanfield2} 
\frac{\overline \Omega\phantom{_0}}{\Omega_0} =   -\int  \frac{\d^3 \r \, \d^3 \r_0}{(1-\zeta) V} v(\r_0+ \r) v(\r_0) n_a\Y_{ab}(\r)\chi_{bc} n_c \, ,
\end{equation} 
For a fixed $\r$, the integration over $\r_0$ is performed over the overlap of two mesoscopic spheres, $r_0<R$ and $|\r_0+\r|<R$, \fig{fig:spheres}. The overlap volume is large for $\r$ of the order of magnitude of the medium correlation length, therefore the $\r_0$-integrated product $v(\r_0+\r) v(\r_0)$ gives rise to the density-density {\em correlation function} of the matrix, 
\begin{equation}\label{def1Gamma}
\Gamma(\r) =  \int \frac{\d\r_0}{V} \,v(\r_0+\r) v(\r_0)  - \zeta^2 \,,
\end{equation}
where $\zeta$ appears as the sample mean of $v(\r)$. The integral from this expression is substituted in \eq{meanfield2}, where the $\zeta^2$ integrated with $\Y_{ab}(\r)$ gives zero by the same reason as above taking into account that the $\r$ integration is performed over the sphere of the radius $2R$. The final expression thus takes the form  
\begin{equation}\label{meanfield_r}
\frac{\overline \Omega\phantom{_0}}{\Omega_0} =  -\frac{1}{1-\zeta} \int \d^3 \r \, n_a\Y_{ab}(\r)\chi_{bc} n_c\,  \Gamma(\r) \, .
\end{equation}

For further analysis, it is convenient to formulate this result in terms of the Fourier-transformed quantities. Straightforward transformation of \eq{def1Gamma} gives 
\begin{equation}\label{GammaF}
\Gamma(\k) =  \frac{1}{V} v(\k) v(-\k)  \,,
\end{equation}
wheret $\Gamma(\k)$ is discontinuous at $\k=0$; its value at this point is zero, which is easy to show by integrating \eq{def1Gamma} over $\r$. The final result in the Fourier domain takes the form  
\begin{equation}\label{meanfield_k}
\frac{\overline \Omega\phantom{_0}}{\Omega_0} = -\frac{1 }{1-\zeta} \int \, \frac{\d^3 k}{(2\pi)^3} \, n_a\Y_{ab}(\k)\chi_{bc} n_c\,  \Gamma(\k) \, ,
\end{equation}
where the elementary dipole field, $\Y_{ab}(\k)$, is defined in \eq{DabLk0}. Note that the tensor $n_a\chi_{bc} n_c$ is constant. Another observation is that the correlation function in \eq{meanfield} is the only quantity that depends on the length of vector $\k$. The radial integration can be performed resulting in the expression 
\begin{equation}\label{meanfield}
\frac{\overline \Omega\phantom{_0}}{\Omega_0} = -\frac{1 }{1-\zeta} \int \, \d^2 \hat k \, n_a\Y_{ab}(\hat k)\chi_{bc} n_c\,  \overline\Gamma(\hat k) \, ,
\end{equation}
where the hats are written to underscore the exclusive dependence of all quantities on the orientation of $\hat k$,  the integration is performed over the full solid angle in $k$-space and 
\be\label{def3Gamma}
\overline\Gamma(\hat k) = \frac{1}{(2\pi)^3}\int_0^\infty dk k^2 \Gamma(\k) \,.
\ee

It is important to notice that the medium's magnetic microarchitechture is represented in the observable frequency shift by only five relevant coefficients. This follows from the observation that the elementary dipole field, \eq{DabLk0} is a trace-free tensor of the second rank build of the unit vector $\hat k$. There is one-to-one correspondence between the components of such tensors and the spherical harmonics, $Y_l^m(\hat k)$, of the order $\ell = 2$ with explicit formulas given in Appendix \ref{ssec:Ylm}. In other words, the expression $n_a\Y_{ab}(\hat k)\chi_{bc} n_c$ in \eq{meanfield} is a linear combination of spherical harmonics $Y_2^m(\hat k)$. Therefore, it performs a projection of all terms in the spherical harmonic expansion of $\overline\Gamma(\hat k)$ onto the subspace with $\ell=2$. By the orthogonality of spherical harmonics, only the projections of the terms with $\ell = 2$ are non-zero. Since $\overline\Gamma(\hat k)$ is real according to \eq{GammaF}, the remaining subspace is five-dimensional spanning the complex-valued coefficients in front of $Y_2^2$, $Y_2^1$ and a real-valued one in front of $Y_2^0$. Note that for the most conventional case of isotropic media, $\overline\Gamma(\hat k)$ is a constant and all coefficients in front of $Y_2^m$ are zero. The mesoscopic sphere thus does not contribute to the average field, in other words, it can be treated as an extension of the classical molecular Lorentz sphere on the mesoscopic scale.

\section{The case of axial symmetry}
\label{sec:cylinders}

The axial symmetry of the medium implies that the susceptibility tensor $\chi_{ab}$ has the eigenvalues $\chi_\perp,\, \chi_\perp,\, \chi_\parallel$ when the third direction is selected along the symmetry axis. When conducting experiments with such media, it makes sense to selects samples of cylindrical shape co-axial with the microscopic symmetry axis, which is assumed throughout this section. Such a setup serves as a (strongly simplified) model of experiments performed with excised nerve segments \cite{Luo2014,Wharton2015}. The aim of this section is to find the frequency shift for an arbitrary orientation of the sample relative to the main field. 

The axial symmetry also implies that $\overline\Gamma(\hat k)$ does not depend on the azimuthal angle, $\varphi$, in the selected reference frame. Therefore, its only relevant component is proportional to $Y_2^0$, 
\be\label{Gamma_relevant}
\overline\Gamma(\hat k) = C_2^0 Y_2^0(\hat k) + \dots \,, 
\ee
where the dotted terms do not contribute to the frequency shift. 

This fact significantly simplifies the calculation of the linear combination $n_a\Y_{ab}(\hat k)\chi_{bc} n_c$ in \eq{meanfield}. In the selected reference frame, the direction of the main field can be chosen as $n = (0,\sin\theta,\cos\theta)^\dagger$ and the product $n_a\Y_{ab}(\hat k)$ can be calculated explicitly using \eq{DabLk0} and omitting all terms that are not contributed by $Y_2^0(\hat k)$. There are only two terms that should be kept according to \eqsand{nyny}{nznz}, which gives $n_a\Y_{ab}(\hat k) = (0,\Y_{22}\sin\theta,\Y_{33}\cos\theta)$. The remaining factor is simply $\chi_{bc}n_c=(0,\chi_\perp\sin\theta,\chi_\parallel\cos\theta)^\dagger$. Taking the product, using the explicit form of the dipole field, \eq{DabLk0} and performing the integration using the normalization of $Y_2^0$ on unity, transform \eq{meanfield} in the following expression for the mean field inside the mesoscopic sphere
\be\label{meanfield_ax_meso}
\frac{\overline \Omega\phantom{_0}}{\Omega_0} = \frac{8\pi^2}{3\sqrt{5\pi}}\frac{C_2^0 }{1-\zeta}\left[(2\chi_\parallel+\chi_\perp)\cos^2\theta - \chi_\perp  \right]  \,.
\ee

The observed frequency shift is found by the addition of the macroscopic contribution defined by $\B^{\rm macro}$ for the overall cylindrical shape with the macroscopically averaged susceptibility tensor, $\zeta\chi_{ab}$ (the first term on the right-hand side of \eq{2fields} and its graphical representation in \fig{fig:sample_decomposition}). A straightforward calculation yields 
\be\label{meanfield_ax_macro}
\frac{\overline \Omega^{\rm macro}\phantom{_0}}{\Omega_0} = \frac{2\pi}{3}\zeta\left[(2\chi_\parallel+\chi_\perp)\cos^2\theta - \chi_\perp  \right]  \,.
\ee

It is instructive to see how the present general approach results in the identical zero frequency shift for a sample made of a bunch of parallel magnetized cylinders with the overall macroscopic cylindrical shape \cite{He2009}. For such media, the coefficient $C_2^0$ can be found for arbitrary structure in the transverse plane (Appendix \ref{ssec:test_cylinders}),  
\be\label{C20cyl}
C_2^0 = -\frac{\sqrt{5\pi}}{4\pi}  \,\zeta(1-\zeta) \,,
\ee
\Eq{meanfield_ax_meso} with this coefficient identically cancels the macroscopic contribution, \eq{meanfield_ax_macro}. The same zero field is obvious when the cylinder-shaped mesoscopic cavity is used instead of the sphere in this special case \cite{He2009,Yablonskiy2014}. 

Note that the coefficient $C_2^0$ is negative in media with elongated microstructure as in the extreme example given by \eq{C20cyl}. In such media, $\Gamma(\r)$ is elongated towards the cigar shape due to stronger correlation along the main axis. The Fourier transform, $\Gamma(\k)$, is correspondingly closer to a disk-shaped form, which, along with the explicit form of $Y_2^0$, \eq{Y20}, leads to negative values of $C_2^0$.

\section{Discussion}
\label{sec:discussion}

The most general result of this study is the mean frequency offset inside the mesoscopic sphere, \eq{meanfield}. When applied to an experiment, this expression has to be combined with frequency shift for the specific macroscopic sample shape calculated according to \eqsand{DBk}{DabLk0}. In addition to the magnetic susceptibility tensor, the medium microarchitecture is represented with only five parameters due to the isomorphism between the elementary dipole field and the spherical harmonics of the order $\ell=2$. 

\subsection{Relevant medium parameters and the fate of the Lorentz cavity}

The present results add to the polemic about the usage of the Lorentz cavity in media with nontrivial microarchitecture \cite{Duyn2014,Yablonskiy2014}. The Lorentz cavity is a simplified way to take into account the near field in \eq{2fields} by subtracting the field of a cavity of a predefined shape \cite{Yablonskiy2014}, the traditional Lorentz sphere for isotropic media \cite{Wolber2000,Durrant2003} or the cylinder for media composed with parallel cylindrical objects \cite{He2009}. One can hypothesize about an ellipsoid as the interpolating shape. Indeed, the aspect ratios and orientation of an ellipsoid are described by five parameters that might be mapped on the five relevant parameters of the medium magnetic microarchitechture. This is not an easy task though and, moreover, it makes little sense: When the near field is found according to \eq{meanfield}, the problem is already solved in the non-simplified manner. The Lorentz cavity construction is thus abandoned for the direct calculation of the magnetic field in agreement with the recent literature \cite{Sukstanskii2014,Wharton2015,Yablonskiy2015,Yablonskiy2017}.  

To my opinion, the most advanced, but still traditional usage of the Lorentz cavity appears in the calculation of the frequency shift in mixtures of isotropic and long parallel  magnetized inclusions with the overall low volume fraction, $\zeta\ll 1$, \cite{Sukstanskii2014,Yablonskiy2015,Yablonskiy2017}. In these calculations, either inclusion type is assigned the own Lorentz cavity, a spherical and a cylindrical one, respectively. This is justified by the property of the correlation function to be proportional to the sum of correlation functions of individual inclusions when $\zeta\ll 1$. Otherwise, the cross-correlations should be taken into account, which breaks the additivity. 

In axially symmetric media, the microarchitecture contributes the single parameter in the frequency shift, the coefficient $C_2^0$ in \eq{meanfield_ax_meso}. Further parameters of interest are the axial and transverse magnetic susceptibilities, $\chi_\parallel$ and $\chi_\perp$. According to \eq{meanfield_ax_meso}, all three parameters cannot be found from measuring the orientation dependence of the frequency shift. The only available are the products of the magnetic susceptibilities with the microarchitecture-defined coefficient $C_2^0$. 

\subsection{Implications for interpretation of previous experiments}

The result expressed in \eq{meanfield_ax_meso} can be considered as a simplified model of experiments in which the signal phase  was measured in excised segments of animal optic nerves as a function of the segment orientation relative to the main field \cite{Luo2014,Wharton2015}. In both experiments, the frequency shift created by the sample in the embedding fluid was used to find the overall sample-averaged magnetic susceptibility that was compared with the frequency shift inside the nerve. Luo et al.\ \cite{Luo2014} interpreted the discovered anisotropy from the microstructural point of view; the sample was considered as consisting of isotropic and cylindrical magnetic susceptibility inclusions, either with a scalar (isotropic) magnetic susceptibility. As discussed above this is a correct quantitative description for inclusions with low volume fraction. Beyond this assumption the decomposition in isotropic and cylindrical inclusion should be considered as an effective representation of the more general result, \eq{meanfield_ax_meso}.  

Wharton and Bowtell \cite{Wharton2015} circumvented the lack of theoretical description for dense media by finding the microstructural contribution as the difference between the measured frequency shift and the traditional approach for the sample with anisotropic magnetic susceptibility. They described the difference as $f_R=A\sin^2\theta + b$, where $A$ represented the microstructure contribution and $b$ was due to both the microstructure and the possible chemical exchange \cite{Zhong2008}. Analysing their framework from the present point of view, the traditional approach coincides with \eq{meanfield_ax_macro} because it takes into account the sphere of Lorentz. Identifying the sample-averaged components of the susceptibility tensor, $\zeta\chi_\parallel$ and $\zeta\chi_\perp$, with $\chi_I+\chi_A$ and $\chi_I-\chi_A/2$ from Ref.\,\cite{Wharton2015}, respectively, results in the following expression for the empirical coefficient $A$:
\be\label{A=}
A = -\frac{4\pi}{\sqrt{5}} \frac{C_2^0}{\zeta(1-\zeta)} \left(\chi_I + \frac{\chi_A}{2} \right) \gamma B_0 \,
\ee
in the cgs system; in SI, the factor $4\pi$ is absorbed in the correspondingly larger numerical values of magnetic susceptibilities. 

The numerical value of this coefficient as found by Wharton and Bowtell results in an essential contribution to the frequency shift. As they showed by simulations in a realistic phantom obtained by translating properties of the optic nerve to the whole human brain, both the quantitative susceptibility mapping (QSM, see Ref.\,\cite{Deistung2017} and references therein) and the susceptibility tensor imaging (STI, see Ref.\,\cite{Li2017} and references therein) cannot quantify the white matter microstructure until the microstructural effects are taken into account. This account, however, requires performing at least diffusion tensor imaging for the determination of local fiber configuration  \cite{Wharton2015}. 

\subsection{Towards realistic model of white matter}

Applications of the obtained results to brain white matter is hindered by essential simplification of the present model. The main one is the location independence of  magnetic susceptibility tensor, $\chi_{ab}$. This allowed the factorization of $\chi_{ab}$ with the remaining terms giving the pure correlation function, $\Gamma$, of the structure, $v(\r)$; for the present discussion the notation can be specified as $\Gamma_{vv}$. In white matter, the susceptibility tensor follows the orientation of myelinated axons \cite{Wharton2012,Sukstanskii2014,Wharton2015,Yablonskiy2015,Yablonskiy2017,Duyn2017,Duyn2017_review}, which are known to have notable orientation dispersion \cite{Leergaard2010,Ronen2014,Veraart2016,Dhital2018}. Extending the present approach to this case requires working with the structure-susceptibility correlation function, $\Gamma_{v\chi}$. While $\Gamma_{vv}$ can be found using, e.g.\ electron microscopy \cite{Raimo2018,Abdollahzadeh2018} finding $\Gamma_{v\chi}$ is more difficult because of the need to assign each point a local magnetic susceptibility, which is invisible in histological images. 

Another problem is the multicompartment structure of white matter. At least two compartment, the intra-axonal and extra-axonal water should be taken into account in any measurement. This is possible in principle, but requires further structuring of the correlation functions in intra- and cross-compartment contributions. It is also worth to note the nontrivial effect of the radially oriented local magnetic susceptibility of myelin sheets \cite{Wharton2012,Sukstanskii2014,Yablonskiy2017}. 

The above problems are not unsolvable, but they require further work to create an adequate theoretical description of the phase contrast in brain white matter and perhaps other anisotropic tissues. Even the oversimplified example considered here shows that the account for microstructure should be essentially more detailed than the simple subtraction of the field of the Lorentz sphere. The account of microstructure is feasible in terms of microstructural correlation functions, which are broadly used in physics to describe the structure of disordered media. 

\section*{Acknowledgments}

I am grateful to Dmitry S.\ Novikov for fruitful discussions. This work was partially supported by the German Research Foundation (DFG), grant KI~1089/6-1.

%%%%%%%%%%%%%%%%%%%%%%%%%%%% Appendix

\section*{Appendices}

\appendix

\section{Spherical harmonics of order $\ell=2$ vs.\ second-rank symmetric trace-free tensor}
\label{ssec:Ylm}

The second-rank symmetric trace-free tensor is the structure appearing in particular in \eq{DabLk0}. Consider a unit three-dimensional vector, ${\bf n}$, that can be specified via its Cartesian components, $n_a$ or the two angles of the spherical co-ordinates, $(n_x,n_y,n_z) = (\sin\theta\cos\varphi, \sin\theta\sin\varphi, \cos\theta)$. Using this relation, it is straightforward to express the spherical harmonics of the second order, $\ell=2$, in terms of $n_a$: 
\begin{align}
Y_2^{-2} &= {1\over 4}\sqrt{{15\over 2\pi}}(n_x-in_y)^2    \\ 
Y_2^{-1} &= {1\over 2}\sqrt{{15\over 2\pi}}(n_x-in_y)n_z    \\ 
Y_2^{0} &= {1\over 4}\sqrt{{5\over \pi}}(2n_z^2-n_x^2-n_y^2)   \label{Y20} \\
Y_2^{1} &= -{1\over 2}\sqrt{{15\over 2\pi}}(n_x+in_y)n_z    \\ 
Y_2^{2} &= {1\over 4}\sqrt{{15\over 2\pi}}(n_x+in_y)^2    
\end{align}
Solving this system gives the inverse transformation, 
\begin{align}
n_x^2-{1\over 3} &= -{2\over 3}\sqrt{{\pi \over 5}} Y_2^{0} + \sqrt{{2\pi\over 15}} \left(Y_2^{ 2} + Y_2^{ -2}\right)   \\ 
n_y^2-{1\over 3} &= -{2\over 3}\sqrt{{\pi \over 5}} Y_2^{0} - \sqrt{{2\pi\over 15}} \left(Y_2^{ 2} + Y_2^{ -2}\right)  \label{nyny} \\  
n_z^2-{1\over 3} &= {4\over 3}\sqrt{{\pi \over 5}} Y_2^{0}  \label{nznz} \\  
n_x n_y &= -i \sqrt{{2\pi\over 15}} \left(Y_2^{ 2} - Y_2^{ -2}\right)  \\  
n_x n_z &= \sqrt{{2\pi\over 15}} \left(Y_2^{ 1} + Y_2^{ -1}\right)  \\ \label{sol_nn=Y} 
n_y n_z &= -i \sqrt{{2\pi\over 15}} \left(Y_2^{ 1} - Y_2^{ -1}\right)   
\end{align}
The quantities on the left-hand sides define all components of the tensor $n_a n_b - \delta_{ab}/3$.

\section{Coefficient $C_2^0$ for the case of parallel cylinders}
\label{ssec:test_cylinders}

The correlation function in this case does not depend on the third co-ordinate, which implies the following form in the Fourier domain:  
\begin{equation}
\Gamma(\k) = 2\pi \, \delta(k_3) \, \Gamma^{(2d)}(k_1,k_2) \, ,
\label{gamma2d} 
\end{equation}
where $\delta(k_z)$ is the Dirac delta-function and $\Gamma^{(2d)}$ the two-dimensional correlation function in the transverse cross-section of the sample. 

It is now straightforward to calculate the coefficient $C_2^0$ in \eq{meanfield_ax_meso}, 
\be\label{C20cyl0}
C_2^0 =\int \frac{dk k^2 d^2 \hat k}{(2\pi)^3} \, 2\pi \, \delta(k_3) \, \Gamma^{(2d)}(k_1,k_2) Y_2^0(\hat k) \,.
\ee
The three-dimensional integration is restored in this expression. It is therefore convenient to substitute $Y_2^0(\hat k)$ with its form in terms of the Cartesian components of $\hat k$, \eq{Y20}, and set $k_3=0$ due to the presence of $\delta(k_3)$. The two-dimensional correlation function is integrated in the transverse plane according to its relation to the variance of the indicator function and the property $v(\r)^2=v(\r)$: 
\begin{equation}\label{intcorr2d}
\int \frac{\d^2 k}{(2\pi)^2} \, \Gamma^{(2d)}(k) \ = \ \Gamma^{(2d)}(r=0) \ = \ \zeta (1- \zeta) \, . 
\end{equation}
This results in \eq{C20cyl}. 

Substitution of this coefficient in the general expression, \eq{meanfield_ax_meso}, results in the following contribution of the mesoscopic sphere  to the frequency shift 
\be\label{meso_cyl}
\frac{\overline \Omega\phantom{_0}}{\Omega_0} = -\frac{2\pi}{3}\zeta \left[(2\chi_\parallel+\chi_\perp)\cos^2\theta - \chi_\perp  \right]  \,. 
\ee
Since this is exactly opposite to the macroscopic contribution, \eq{meanfield_ax_macro}, the whole cylindrical sample does not result in any frequency shift.

%%%%%%%%%%%%%%%%%%%%%%%%%%%%%%%%%%%%%%%%%%%%
\end{document}